\documentclass[letterpaper,twocolumn,american,amsmath,amssymb,preprintnumbers,showpacs,showkeys,aps,prl,superscriptaddress]{revtex4-2}
\usepackage{mathptmx}
\usepackage[T1]{fontenc}
\usepackage[latin9]{inputenc}
\usepackage{color}
\usepackage{babel}
\usepackage{verbatim}
\usepackage{float}
\usepackage{amsmath}
\usepackage{amssymb}
\usepackage{graphicx}
\usepackage[unicode=true,pdfusetitle,
 bookmarks=true,bookmarksnumbered=false,bookmarksopen=false,
 breaklinks=false,pdfborder={0 0 1},backref=false,colorlinks=true]
 {hyperref}

\makeatletter

\usepackage{babel}

\makeatother

\begin{document}
\title{Probing non-Hermitian phase transitions in curved space via quench
dynamics}
\author{Ygor Pará}
\affiliation{Departamento de Física Teórica e Experimental, Universidade Federal
do Rio Grande do Norte, 59072-970 Natal-RN, Brazil}
\email{ygpara@fisica.ufrn.br}

\author{Giandomenico Palumbo}
\affiliation{School of Theoretical Physics, Dublin Institute for Advanced Studies,
10 Burlington Road, Dublin 4, Ireland}
\email{giandomenico.palumbo@gmail.com}

\author{Tommaso Macrì}
\affiliation{Departamento de Física Teórica e Experimental, Universidade Federal
do Rio Grande do Norte, 59072-970 Natal-RN, Brazil}
\affiliation{International Institute of Physics, Universidade Federal do Rio Grande
do Norte, 59078-400 Natal-RN, Brazil}
\email{macri@fisica.ufrn.br}

\begin{abstract}
Non-Hermitian Hamiltonians are relevant to describe the features of
a broad class of physical phenomena, ranging from photonics and atomic
and molecular systems, to nuclear physics and mesoscopic electronic
systems. An important question relies on the understanding of the
influence of curved background on the static and dynamical properties
of non-Hermitian systems. In this work, we study the interplay of
geometry and non-Hermitian dynamics by unveiling the existence of
curvature-dependent non-Hermitian phase transitions. We investigate
a prototypical model of Dirac fermions on a sphere with an imaginary
mass term. This exactly-solvable model admits an infinite set of curvature-dependent
pseudo-Landau levels. We characterize these phases by computing an
order parameter given by the pseudo-magnetization and, independently,
the non-Hermitian fidelity susceptibility. Finally, we probe the non-Hermitian
phase transitions by computing the (generalized) Loschmidt echo and
the dynamical fidelity after a quantum quench of the imaginary mass
and find singularities in correspondence of exceptional radii of the
sphere. 
\end{abstract}
\maketitle

\section{\label{sec:introduction}Introduction}

Non-Hermitian (NH) topological systems have become a very active and
prolific research field in recent years \citep{Sato,Sato2,Ueda2,Kunst,Torres,Lee2,Fu2,Szameit2,yamamoto2019,yoshida2019}.
This is due to the their novel and peculiar physical features such
as the NH skin effect \citep{Sato4,Thomale,Longhi2,zhang2020.3} and
the lacking of the conventional bulk-edge correspondence \citep{Kunst3,Wang,Zhang,Slager,Takane},
as well as to the plethora of experimental platforms where these systems
can simulated and tested. A special class of non-Hermitian phases
is defined by the $\mathcal{PT}$ (parity and time reversal (TR))
symmetry, where the spectrum is completely real even with the absence
of the Hermitian condition \citep{Bender,Rui}. This symmetry is important
also because NH phase transitions are deeply related to the existence
of the exceptional points (EPs) \citep{Kunst2,Sato3,Hatsugai,Gong,Smith2},
which are associated to the critical phase transitions between $\mathcal{PT}$-symmetric
and $\mathcal{PT}$-broken phases. These special points are characterized
by topological invariants \citep{Nori2}. Moreover, the presence of
EPs can be tested at dynamical level by employing genuine information-theoretic
quantifiers such as (generalized) fidelities and Loschmidt echos \citep{tzeng2020hunting,Kou,Ueda1,longhi,Jafari}.
Notice, these powerful theoretical tools have been largely used in
the study of Hermitian phase transitions \citep{Gorin_2006,Mera2,Grandi,Zanardi3,Vieira,Capponi,Gu,PhysRevA.94.010102,Goussev2016,Zanardi1,Zanardi2,Moore,Jafari,Mera1,pires2020,macieszczak2019}.

However, differently from Hermitian systems, the behavior of NH phases
is still poorly understood on curved background. Curved space for
emergent relativistic fermions in condensed matter physics can be
associated, for instance, to the physical curvature of $1$D, $2$D
and 3D materials, to effective geometry created from the synthetic
matter or to formal geometric approaches employed to study the quantum
thermal effects in topological phases \citep{Cortijo,Boada_2011,Ryu,Stone2,Hughes,Palumbo1,Lambiase,Palumbo2,Angelakis,Landsteiner,Palumbo3,Celi,Ojanen,Smith,Palumbo4,Nissinen2,Nissinen3,Farjami}.
Importantly, it has also been shown that curved background can give
rise to genuine Hermitian phase transitions \citep{Ishikawa,Ortix,Flachi,Ghee}.

The main goal of our work is to unveil the existence of curvature-dependent
non-Hermitian phase transitions. These non-Hermitian phases are characterized
by EPs which depend on the geometry of the system. In particular,
these \emph{geometric} EPs exist only on curved space and in general
are hard to identify and describe due to the lack of translation symmetry.
For this reason, we will present and study a prototypical analytical
solvable model in two dimensions given by Dirac fermions on the sphere.
In the Hermitian case, this model arises in fullerene \citep{2002hep.th...12134A,Abrikosov:2001nj,Kolesnikov2006,vozmediano,VOZMEDIANO2010109,Pincak_2005,PachosStone}
and as a surface model of three-dimensional topological insulators
\citep{doi:10.7566/JPSJ.82.074712,fukui,Lee}. Importantly, experimental
realizations with superconducting resonators \citep{PhysRevLett.115.026801}
and topological nanoparticles \citep{Siroki2016,Rider2020} recently
appeared. We will induce NH phases by introducing a complex Dirac
mass, which can be implemented by introducing on-site gain and/or
loss similarly to the flat case \citep{Wang,Szameit,Nori,Leykam,Fu2,Zhao}.
This situation render the Hamiltonian always intrinsically non-Hermitian
\citep{xue2020nonhermitian} inducing exceptional points in the energy
spectrum.

Firstly, we will analytically solve the model by showing the emergence
of NH pseudo-Landau levels and their discrete symmetries. Notice,
these pseudo-Landau levels are related to the intrinsic curvature
of the sphere and are the analog of pseudo-Landau levels induced in
flat space through strain \citep{Schomerus,deJuan2,Franz,Vojta,Grushin2,Salerno,Stone,Nissinen}.

Secondly, we will present the infinite sequence of geometric EPs,
which will directly depend on the radius (curvature) of the sphere
and relate the static properties of the spectrum through a pseudo-magnetization
induced by the imaginary Dirac mass.

Finally, we will employ quench dynamics to characterize these curvature-dependent
NH phase transitions. In particular, we will show that the NH versions
of the Loschmidt echo and fidelity \citep{longhi} are able to distinguish
between the different phases of the model. The existence of non-analytic
points in the fidelity will represent the clear signature of the presence
of the geometric EPs. This will be performed by quenching the imaginary
mass term and tuning the radius of the sphere to different values.

\section{Model}

We consider the dynamics of relativistic fermions constrained to the
surface of a sphere of radius $R$. We will induce Non-Hermitian phases
by introducing an imaginary Dirac mass, which can be implemented by
introducing on-site gain and/or loss similarly to the flat case \citep{Wang,Szameit,Nori,Leykam,Fu2,Zhao}.
The Dirac equation in 2+1 dimensions for a fermion of \emph{complex}
mass $\mu=M+i\Gamma$ (with both $M$ and $\Gamma$ real dimensionfull
constants) in a curved background is described by a two-component
spinor $\Psi(\mathbf{x})$ reads \citep{Wald:1984rg,Weinberg,Pollock:2010zz,Collas_2019}
\begin{equation}
\left(i\hbar\bar{\gamma}^{\nu}\nabla_{\nu}+\mu\,c\right)\Psi(\mathbf{x})=0.\label{eq:dirac}
\end{equation}
 The space-time coordinates are labeled by $\mathbf{x}=(c\,t,\theta,\phi)$,
the covariant derivative is given by $\nabla_{\nu}=\partial_{\nu}+\frac{1}{8}\omega_{\nu bc}\left[\gamma^{b},\gamma^{c}\right]$.
The spin connection $\omega_{\mu b}^{d}=e_{\nu}^{\ \ d}\left(e_{\ \ b}^{\sigma}\Gamma_{\ \ \sigma\mu}^{\nu}+\partial_{\mu}e_{\ \ b}^{\nu}\right)$
is written in terms of the metric connection $\Gamma_{\ \ \sigma\mu}^{\nu}=\frac{1}{2}g^{\beta\nu}\left(\partial_{\mu}g_{\sigma\beta}-\partial_{\beta}g_{\sigma\mu}+\partial_{\sigma}g_{\mu\beta}\right).$
We are using $a,b,\ldots$ for local frame indices and $\mu,\nu,\ldots$
for coordinate indices. The metric of this space is $g_{\mu\nu}=\text{diag}\left(1,-R^{2},-R^{2}\sin^{2}\theta\right)$,
with determinant $g=R^{4}\sin^{2}\theta$. The \emph{vielbein} is
given by the definition $g_{\mu\nu}=\eta_{ab}e_{\mu}^{\ \ a}e_{\nu}^{\ \ b},$
where $\eta_{ab}=\text{diag}\left(1,-1,-1\right)$. We also defined
the inverse \emph{vielbein} to satisfy $e_{\mu}^{\ \ a}e_{\ \ b}^{\mu}=\delta_{b}^{a}.$
We write the gamma matrices as $\bar{\gamma}^{\mu}=e_{\ \ a}^{\mu}\gamma^{a}$,
where we choose $\gamma^{0}=\sigma_{z},$ $\gamma^{1}=-i\sigma_{x},$
$\gamma^{2}=-i\sigma_{y},$ consistently with the anti-commutation
relations $\left\{ \bar{\gamma}^{\mu},\bar{\gamma}^{\nu}\right\} =2g^{\mu\nu}$
\citep{Arminjon_2014}.

The massless case represents the effective low-energy model for the
some spherical condensed matter systems such as fullerene and the
boundary states of spherical topological insulators \citep{2002hep.th...12134A,Abrikosov:2001nj,doi:10.7566/JPSJ.82.074712,Kolesnikov2006,vozmediano,VOZMEDIANO2010109,Pincak_2005,PhysRevLett.69.172,fukui,Hasebe,Lee,Hsiao}.
It has also been employed in the prediction of some spectral properties
related to the number of zero modes, which are theoretically predicted
by the Atiyah-Singer index theorem \citep{PachosStone, PhysRevLett.115.026801}.
In the context of three-dimensional topological insulators, the model
in Eq.(\ref{eq:dirac}) displays several many-body phases in presence
of attractive and in repulsive interactions \citep{PhysRevLett.115.017001}.

By separating the time component from the spatial ones in Eq. (\ref{eq:dirac}),
we can write a Dirac Hamiltonian in the form 
\begin{equation}
\hat{H}_{\mu}=\left(\begin{array}{cc}
-\mu\,c^{2} & \frac{\hbar c}{R}\hat{h}^{-}\\
\frac{\hbar c}{R}\hat{h}^{+} & \mu\,c^{2}
\end{array}\right),\label{eq:H}
\end{equation}
with $\hat{h}^{\pm}=\pm\left(\partial_{\theta}+\frac{\cot\theta}{2}\right)+\frac{i}{\sin\theta}\partial_{\varphi}.$
It is important to note that due to the presence of the NH mass term
$\hat{\Sigma}_{z}=-\mu c^{2}\sigma_{z}$, Eq.(\ref{eq:H}) is a NH
Hamiltonian, i.e. $\hat{H}_{\mu}\neq\hat{H}_{\mu}^{\dagger}$. The
eigenvalues $E_{\mu,R}^{\left(n,m,\lambda\right)}$ of $H_{\mu}$
can be obtained by squaring the Hamiltonian $\hat{H}_{\mu}$ of Eq.(\ref{eq:H})
\begin{equation}
E_{\mu,R}^{\left(n,m,\lambda\right)}=\lambda\sqrt{\frac{\hbar^{2}c^{2}}{R^{2}}\Omega_{nm}^{2}+\mu^{2}c^{4}},\label{eq:energy}
\end{equation}
where $\lambda=\pm1$ and $\Omega_{nm}=n+\left|m\right|+1/2$.

\begin{figure}[H]
\centering{}\includegraphics[width=1\columnwidth]{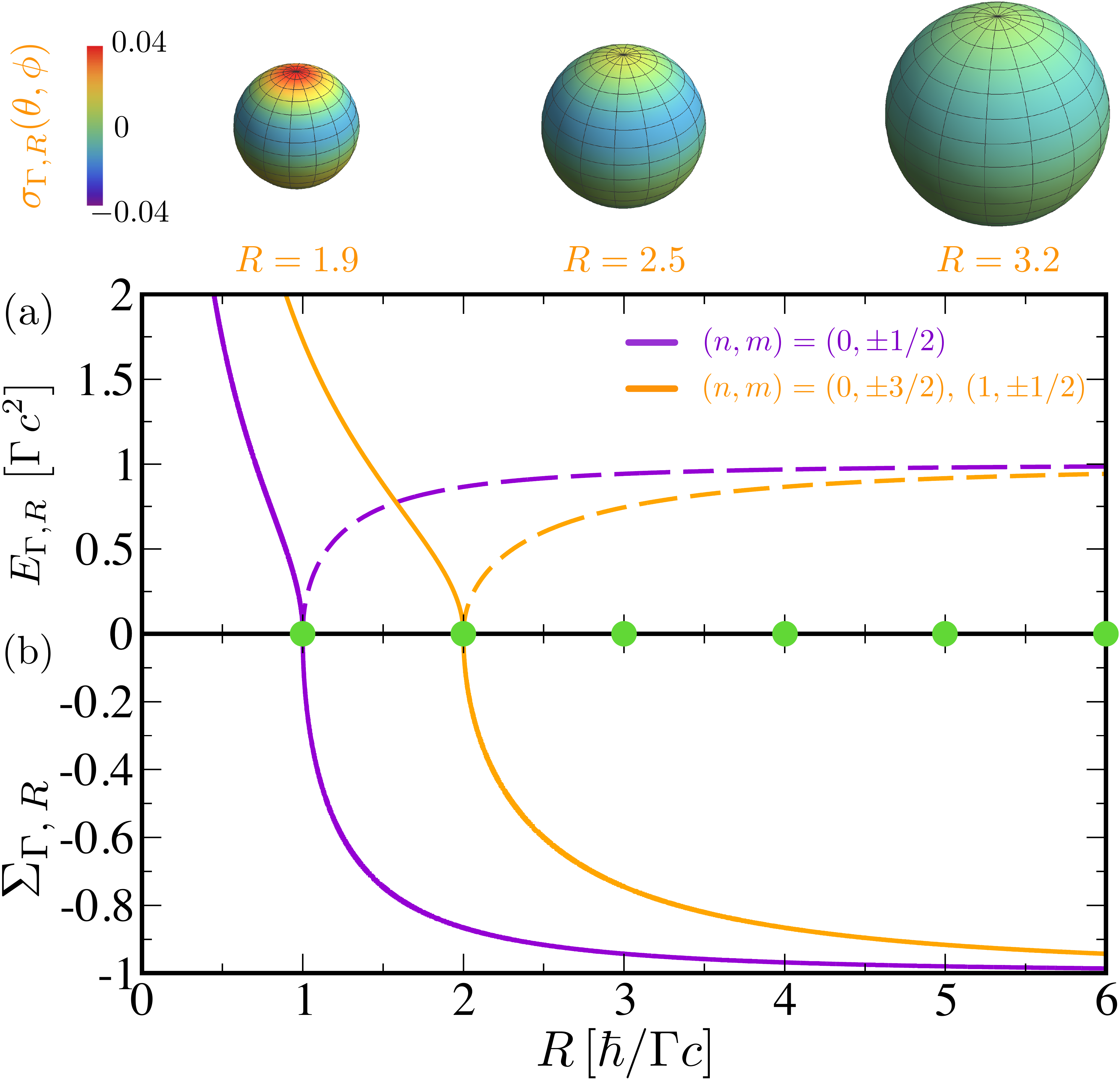} \caption{Eigenvalues and NH magnetization for the eigenstates of Eq.(\ref{eq:dirac})
as a function of the radius $R$ of the sphere. (a) $E_{\Gamma,R}$
for $\lambda=1$ in units of $\Gamma c^{2}$ for the the first two
pLLs with $(n,m)=(0,\pm1/2)$ (purple) and $(1,\pm1/2)$ (orange)
in the TR-invariant region (real, full lines) and in the TR-broken
phase (imaginary, dashed) according to Eq.(\ref{eq:energy}). At the
EPs (green circles) the eigenvalues have vanishing energy, and eigenstates
with opposite signs coalesce with degeneracy $d=4\,\Omega_{nm}$.
(b) $\Sigma_{\Gamma,R}$ from Eq.(\ref{eq:magnetization}) for $\lambda=1$
and for the same pLLs of (a). At the EPs the spontaneous breaking
of TR symmetry leads to a finite magnetization which asymptotically
reaches the value $\Sigma_{\Gamma,\infty}=-1$. In the vicinity of
the EPs $R\gtrsim R^{*}$, magnetization increases $\propto-\lambda(R-R^{*})^{1/2}$.
The energies and magnetizations for $\lambda=-1$ have opposite signs
of $E_{\Gamma,R}$ and $\Sigma_{\Gamma,R}$. Top. Local magnetization
$\sigma_{\Gamma,R}(\theta,\phi)$ across the EP at $R^{*}=2$ for
$(n,m,\lambda)=(1,1/2,1)$. For $R<R^{*}$, $\sigma_{\Gamma,R}(\theta,\phi)=-\sigma_{\Gamma,R}(\pi-\theta,\phi)$.
For $R>R^{*}$, $\sigma_{\Gamma,R}(\theta,\phi)$ varies over a smaller
range, and its integral is negative.}
\label{fig:fig1} 
\end{figure}

The indices $n$, $m$ are respectively a positive integer and a half-integer,
which label the pseudo-Landau levels (pLLs) of the model \citep{GREITER201833,Ilan_2019,PhysRevB.94.035105, Sanchez}.
In the following we will focus on the case where the mass is purely
imaginary, i.e. $\mu\equiv i\,\Gamma$. For this choice, an infinite
set of EPs occur when $E_{\Gamma,R}^{\left(n,m,\lambda\right)}=0$,
i.e. $R\,\Gamma c^{2}=\hbar\,c\,\Omega_{nm}$ with degeneracy $d=4\,\Omega_{nm}.$
At the EPs a pair of opposite-sign eigenvalues, connected by a square
root branch point, coalesce \citep{PhysRevLett.80.5243,Kato1995}.
The evolution of the eigenvalues for the first pLLs are shown in Fig.(\ref{fig:fig1})a
as a function of the sphere radius $R$. The EPs are located at $R=R^{*}$,
with $R^{*}=1,\,2$ respectively. For $0<R<R^{*}$ the eigenvalues
are real, whereas for $R>R^{*}$ they are imaginary.

We notice that the two-component spinor eigenvectors $\phi_{\Gamma,R}^{\left(n,m,\lambda\right)}\left(\theta,\varphi\right)$
of $\hat{H}_{\Gamma}$ have a nonvanishing scalar product 
\begin{align}
\int_{0}^{\pi}\int_{0}^{2\pi}\left(\phi_{\Gamma,R}^{\left(n_{1},m_{1},\lambda_{1}\right)}\right)^{\dagger}\phi_{\Gamma,R}^{\left(n_{2},m_{2},\lambda_{2}\right)}\sqrt{|g|}d\varphi\,d\theta\, & =\nonumber \\
C_{\lambda_{1},\lambda_{2}}\delta_{n_{1},n_{2}}\delta_{m_{1},m_{2}},
\end{align}
with $C_{\lambda_{1},\lambda_{2}}\neq0$ for any nonzero value of
$\Gamma$, and therefore they form a non-orthogonal basis \citep{PhysRevB.83.115129,PhysRevLett.51.605}.
The prefactor of $\phi_{\Gamma,R}^{\left(n,m,\lambda\right)}\left(\theta,\varphi\right)$
is fixed by the requirement that the normalization constant is unitary.
See the Appendix \ref{sec:Spectrum} for a detailed derivation of
the spectrum.

\subsection{Discrete symmetries}

The time reversal (TR) operator is described by $\mathcal{T}=-i\sigma_{y}K$,
where $K$ is the usual complex conjugation. The massless Dirac Hamiltonian
$\hat{H}_{0}$ in Eq.(\ref{eq:H}) with $\Gamma=0$ is Hermitian,
and it is simultaneously invariant under the action of both the parity
and TR symmetry \citep{PhysRevLett.115.017001,PhysRevD.79.024020,fukui,Boada_2011,CAMPORESI19961}.
Any finite (real or complex) mass breaks parity symmetry. Conversely,
the Hamiltonian $\hat{H}_{\Gamma}$ is invariant under TR symmetry.
We also notice that, in our relativistic formulation, TR acts in a
similar fashion as a combined $\mathcal{PT}$-symmetry for a single
spin system with no center-of-mass dynamics \citep{Bender}. The TR
operator $\mathcal{T}$ acts on the eigenstates $\phi_{\Gamma,R}^{\left(n,m,\lambda\right)}$
of $\hat{H}_{\Gamma}$ as follows 
\begin{equation}
\mathcal{T}\phi_{\Gamma,R}^{\left(n,m,\lambda\right)}=\begin{cases}
(-1)^{m-\frac{1}{2}}\phi_{\Gamma,R}^{\left(n,-m,\lambda\right)} & \text{for }\hbar\,c\,\Omega_{nm}>R\,\Gamma c^{2}\\
(-1)^{m-\frac{1}{2}}\phi_{\Gamma,R}^{\left(n,-m,-\lambda\right)} & \text{for }\hbar\,c\,\Omega_{nm}<R\,\Gamma c^{2}
\end{cases}.
\end{equation}
Importantly, upon crossing any EPs, such that $R\,\Gamma c^{2}>\hbar\,c\,\Omega_{nm}$
the effect of $\mathcal{T}$ is to map an eigenstate $\phi_{\Gamma,R}^{\left(n,m,\lambda\right)}$
into an eigenstate $\phi_{\Gamma,R}^{\left(n,-m,-\lambda\right)}$
with opposite energy ($\lambda\rightarrow-\lambda$). Therefore an
adiabatic variation of the radius (geometry) or the imaginary mass
(gain and/or loss) across an EP \emph{spontaneously} breaks TR symmetry.
The TR symmetry due to Kramers' degeneracy, the real eigenvalues are
always paired and can be split into two complex eigenvalue pairs $(E,\ E^{\ast})$.

The parity operator, equivalent to space reflection in $2+1$ dimensions,
can be written as $\mathcal{P}=\sigma_{x}P_{\theta}$, where $P_{\theta}$
takes $\theta\rightarrow\pi-\theta$. But, the role of the proper
parity operator in the model is played by the operator $\mathcal{P}_{0}$
which maps $\psi\left(\theta,\varphi\right)\rightarrow i\sigma_{z}\psi\left(\pi-\theta,-\varphi\right)$.
This operator corresponds to a rotation around the $z$-direction
in the spin degree of freedom and to a reflection with respect to
the origin in the spatial coordinates. Combined with the TR operator
satisfies indeed $(\mathcal{T}\mathcal{P}_{0})^{2}=1$. The proof
makes use of the properties of Jacobi polynomials and algebraic manipulations
of the spinor under the transformations above. Also $\mathcal{T}\mathcal{P}_{0}$
signals the presence of the $\mathcal{PT}\text{-transition}$, as
in the symmetric phase the eigenstates of the Hamiltonian are also
eigenstates of $\mathcal{T}\mathcal{P}_{0}$. Instead, in the broken
phases the eigenstates are no more eigenstates of $\mathcal{T}\mathcal{P}_{0}$,
in a analogy to the usual $\mathcal{PT}\text{-transition}$.

\subsection{Biorthogonal basis.}

Several properties of a NH system can be obtained by the use of the
biorthogonal basis. 
 The eigenstates of $\hat{H}_{\Gamma}^{\dagger}$ satisfy the eigenvalue
equation $\hat{H}_{\Gamma}^{\dagger}\chi_{nm}^{\left(\lambda\right)}=\Xi_{nm}^{\left(\lambda\right)}\chi_{nm}^{\left(\lambda\right)}.$
Note that the Hamiltonians $\hat{H}_{\Gamma}$ and $\hat{H}_{\Gamma}^{\dagger}$
differ only by the sign of $\Gamma$. The eigenvalues are related
by $\Xi_{\Gamma,R}^{\left(n,m,\lambda\right)}=\left[E_{\Gamma,R}^{\left(n,m,\lambda\right)}\right]^{\ast}$.
The eigenvectors of $\hat{H}_{\Gamma}^{\dagger}$ are given by lengthy
expressions reported in Appendix \ref{sec:Biorthogonal}. The set
of the eigenstates of $\hat{H}_{\Gamma}$ and $\hat{H}_{\Gamma}^{\dagger}$
form a biorthogonal basis with a metric-dependent scalar product defined
as \citep{Brody_2013,castroalvaredo2009,Kleefeld1,2016,Moiseyev_2009}
\begin{align}
\int_{0}^{\pi}\int_{0}^{2\pi}\left[\chi_{\Gamma,R}^{\left(n_{1},m_{1},\lambda_{1}\right)}\right]^{\dagger}\psi_{\Gamma,R}^{\left(n_{2},m_{2},\lambda_{2}\right)}\sqrt{|g|}d\varphi\,d\theta & =\nonumber \\
=\delta_{n_{1},n_{2}}\delta_{m_{1},m_{2}} & \delta_{\lambda_{1},\lambda_{2}}.\label{eq:bi-ortho-scalar-product}
\end{align}

\section{Results}

\subsection{Magnetic phases}

Based on the analysis of TR symmetry, and the analogy with usual magnetic
systems, we expect the system to develop a spontaneous magnetization
across an EP \citep{Wei2017,Streubel_2016}. Given a general spinor
$\Phi=\left(\phi_{\uparrow}\ \ \phi_{\downarrow}\right)^{\text{T}}$
we define a NH magnetization as $\Sigma_{\Gamma,R}=\int_{0}^{2\pi}\int_{0}^{\pi}\sigma_{\Gamma,R}(\theta,\phi)\sqrt{|g|}d\theta d\varphi$,
where $\sigma_{\Gamma,R}(\theta,\phi)=(\phi_{\uparrow}^{\ast}\phi_{\uparrow}-\phi_{\downarrow}^{\ast}\phi_{\downarrow})$
is the local magnetization (see top panel of Fig.(\ref{fig:fig1}))
\citep{PhysRevLett.115.017001}. For an eigenstate $\phi_{\Gamma,R}^{\left(n,m,\lambda\right)}$
the NH magnetization can be computed analytically 
\begin{align}
\Sigma_{\Gamma,R}^{\left(n,m,\lambda\right)} & =\frac{\frac{\hbar^{2}c^{2}}{R^{2}}\Omega_{nm}^{2}-\left|E_{\Gamma,R}^{\left(n,m,\lambda\right)}+i\Gamma c^{2}\right|^{2}}{\frac{\hbar^{2}c^{2}}{R^{2}}\Omega_{nm}^{2}+\left|E_{\Gamma,R}^{\left(n,m,\lambda\right)}+i\Gamma c^{2}\right|^{2}}.\label{eq:magnetization}
\end{align}
In Fig.(\ref{fig:fig1})b we plot the magnetization for the pLL as
a function of $R$. For $R<R^{*}$ the magnetization vanishes. At
the critical radius $R=R^{*}$ the system develops a spontaneous magnetization.
In the vicinity of an EP, with $R\gtrsim\hbar\,c\,\Omega_{nm}/\Gamma c^{2}$,
we can expand Eq.(\ref{eq:magnetization}) to obtain 
\begin{align}
\left.\Sigma_{\Gamma,R}^{\left(n,m,\lambda\right)}\right|_{R\gtrsim R^{*}} & \approx-\lambda\sqrt{2}\left(\frac{R-R^{*}}{R^{*}}\right)^{\frac{1}{2}},\label{eq:expansion}
\end{align}
where $R^{*}=\hbar\,\Omega_{nm}/\Gamma\,c$ is the critical radius.
Therefore the magnetization signals the presence a second-order phase
transition with a critical exponent $\beta=1/2$, analogous to the
mean-field critical exponent of the (Hermitian) Ising transition.
Asymptotically, for $R\rightarrow\infty$ the magnetization saturates
to $\Sigma_{\Gamma,\infty}=-\lambda$ depending on the sign of the
energy eigenvalue. We emphasize that the NH magnetization $\Sigma_{\Gamma,R}^{\left(n,m,\lambda\right)}$
defined in Eq.(\ref{eq:magnetization}) relies on the use of the non-orthogonal
basis $\{\phi_{\Gamma,R}^{\left(n,m,\lambda\right)}\left(\theta,\varphi\right)\}$.
Note that if we define NH magnetization using the biorthogonal basis
and redefining appropriately the Pauli matrices \citep{Brody_2013,castroalvaredo2009},
the result is not sensitive to the EPs. To probe such NH phase transitions
in curved space we will use other measurable quantities as described
thoroughly in the following sections. 

\subsection{Fidelity susceptibility}

\begin{figure}[H]
\includegraphics[width=1\columnwidth]{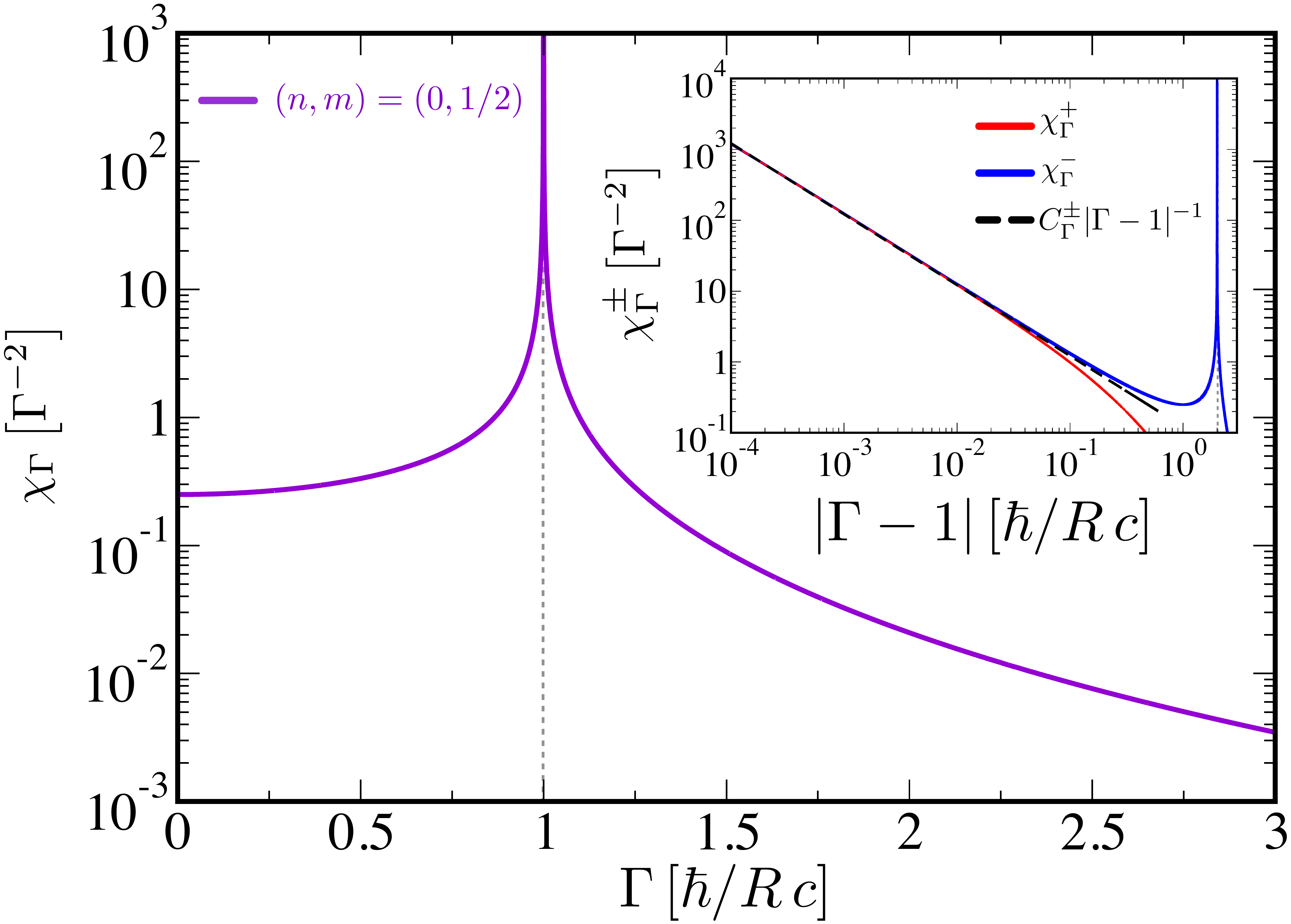} \caption{Fidelity susceptibility $\chi_{\Gamma}$ given by Eq. (\ref{eq:suscept})
of the lowest pLL for $n=0$ and $m=1/2$, as function of the mass
$\Gamma$. Susceptibility $\chi_{\Gamma}$ diverges at the EP $\Gamma^{*}=\hbar\,\Omega_{0,1/2}/R\,c$.
$\chi_{\Gamma}$ is a symmetric function of $\Gamma$, i.e. $\chi_{\Gamma}=\chi_{-\Gamma}$.
Scaling of $\chi_{\Gamma}$ approaching the EP from the right (red
line) and from the left (blue). Both susceptibilities were fitted
with the function $\chi_{\Gamma}^{\pm}=C_{\Gamma}^{\pm}|\Gamma-\Gamma^{*}|^{-1}$,
with $C_{\Gamma}^{\pm}=0.122$. The second peak of $\chi_{\Gamma}^{-}$
corresponds to the EP at $\Gamma=-\hbar\,\Omega_{0,1/2}/R\,c$ outside
the scaling region of the inset.}
\label{fig:fig2} 
\end{figure}

A complementary approach to detect phase transitions \emph{without}
a previous knowledge of an order parameter is via the calculation
of the fidelity susceptibility \citep{Gorin_2006,Mera2,Zanardi3,Vieira,Capponi,Gu}.
This approach has been recently generalized to the study of NH models
\citep{tzeng2020hunting,Kou,Ueda1}. Given two states $|\psi_{\Gamma_{1,2}}\rangle$,
we define the NH fidelity is defined as \citep{longhi} 
\begin{equation}
\mathcal{F}_{R}\{|\psi_{\Gamma_{1}}\rangle,|\psi_{\Gamma_{2}}\rangle\}=\frac{\left|\langle\psi_{\Gamma_{1}}|\psi_{\Gamma_{2}}\rangle\right|^{2}}{\langle\psi_{\Gamma_{1}}|\psi_{\Gamma_{1}}\rangle\langle\psi_{\Gamma_{2}}|\psi_{\Gamma_{2}}\rangle}.\label{eq:fidelity}
\end{equation}
Notice that the scalar product $\langle\psi_{\Gamma_{1,2}}|\psi_{\Gamma_{1,2}}\rangle$
has to be computed at fixed $\lambda$ and radius $R$, whereas other
parameters (such as the mass $\Gamma$ or time $t$) can vary between
the two states. The denominator in Eq.(\ref{eq:fidelity}) bounds
the fidelity, $0\le\mathcal{F}_{R}\le1$. Given two states $|\psi_{\Gamma}\rangle$
and $|\psi_{\Gamma+\delta\Gamma}\rangle$, differing from an infinitesimal
variation of the mass $\Gamma$ one can define the nonorthogonal \emph{static}
fidelity susceptibility as 
\begin{equation}
\chi_{\Gamma}=\lim_{\delta\Gamma\rightarrow0}\frac{-\ln\left|\mathcal{F}_{R}\{|\psi_{\Gamma}\rangle,|\psi_{\Gamma+\delta\Gamma}\rangle\}\right|}{(\delta\Gamma)^{2}}.\label{eq:suscept}
\end{equation}
In Fig.(\ref{fig:fig2}) we plot $\chi_{\Gamma}$ for the lowest pLL
with $(n,m,\lambda)=(0,1/2,1)$. The susceptibility satisfies $\chi_{\Gamma}=\chi_{-\Gamma}$
and diverges at the EP when $\Gamma=\hbar\,\Omega_{n,m}/R\,c.$ In
the inset of Fig.(\ref{fig:fig2}) we compute the scaling properties
of $\chi_{\Gamma}$ around the EP at $\Gamma^{*}=1$. Fitting the
right ($\chi_{\Gamma}^{+}$) and the left ($\chi_{\Gamma}^{-}$) susceptibilities
with the scaling functions $C_{\Gamma}^{\pm}|\Gamma-1|^{\gamma}$
we find that $\gamma=1$ with the same prefactor $C_{\Gamma}^{\pm}=0.12$.
Interestingly, we verified that the fidelity susceptibility has a
qualitatively similar behavior adopting biorthogonal basis states.
However the critical exponent $\gamma$ \emph{is} dependent on the
choice of the basis states. 

The fidelity susceptibility in the biorthogonal basis is defined as
\begin{equation}
\chi_{\Gamma}^{\text{(B)}}=\lim_{\delta\Gamma\rightarrow0}\frac{-2\ln\left|F_{\Gamma}^{(B)}\right|}{(\delta\Gamma)^{2}},\label{eq:suscept_b}
\end{equation}
where $F_{\Gamma}^{(B)}=(\langle\chi_{\Gamma+\delta\Gamma,R}^{\left(n,m,\lambda\right)}|\psi_{\Gamma,R}^{\left(n,m,\lambda\right)}\rangle\langle\chi_{\Gamma,R}^{\left(n,m,\lambda\right)}|\psi_{\Gamma+\delta\Gamma,R}^{\left(n,m,\lambda\right)}\rangle)^{1/2}$
is the generalized Uhlmann fidelity computed on the eigenstates of
$\hat{H}_{\Gamma}$ and $\hat{H}_{\Gamma}^{\dagger}$ with the help
of the biorthogonal basis \citep{Gorin_2006}.

\begin{figure}[H]
\begin{centering}
\includegraphics[width=1\columnwidth]{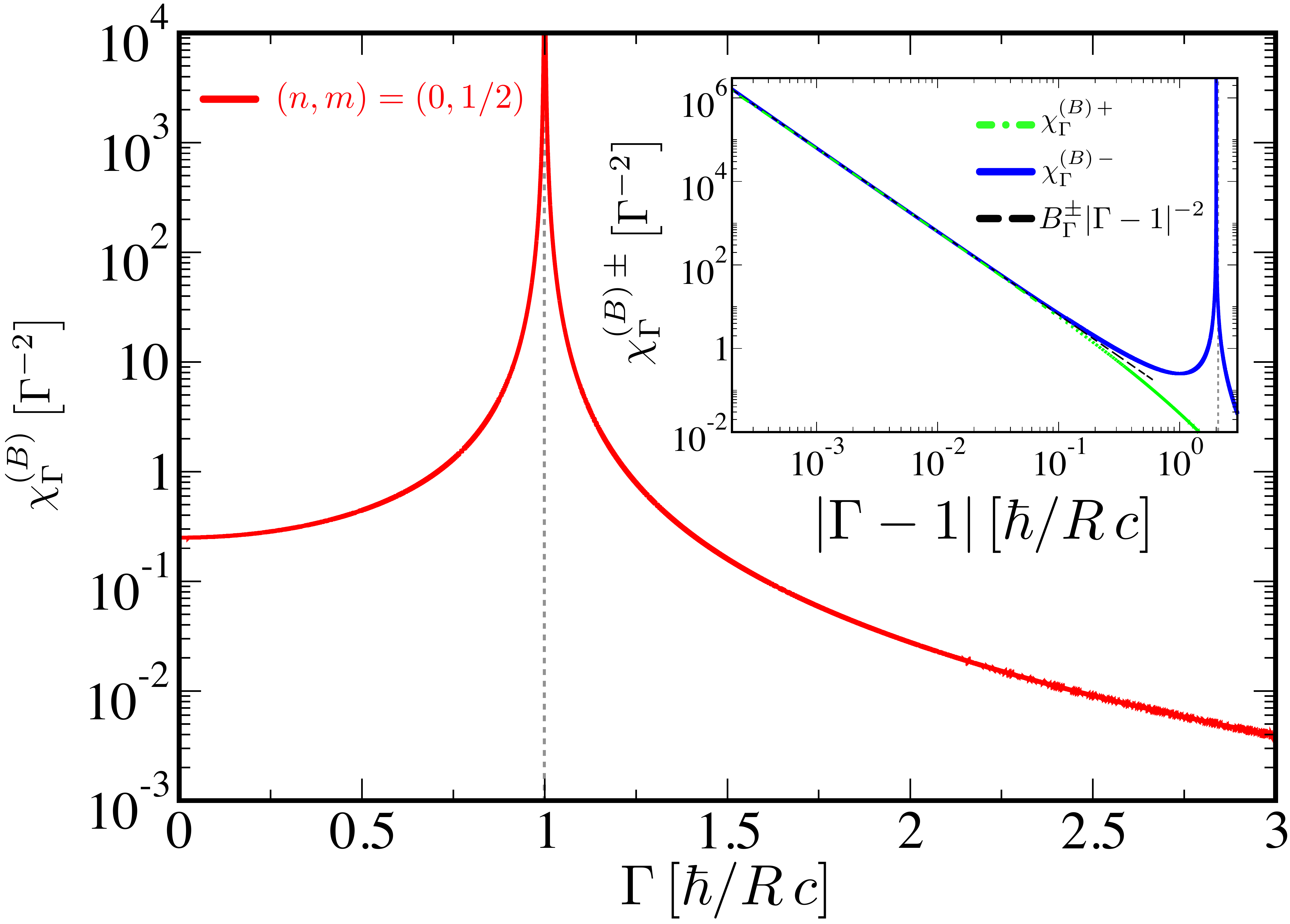} 
\par\end{centering}
\caption{Biorthogonal fidelity susceptibility $\chi_{\Gamma}^{(B)}$ given
by Eq. (\ref{eq:suscept_b}) of the lowest pLL for $n=0$ and $m=1/2$,
as function of the mass $\Gamma$. Susceptibility $\chi_{\Gamma}$
diverges at the EP $\Gamma^{*}=\hbar\,\Omega_{0,1/2}/R\,c$. $\chi_{\Gamma}^{(B)}$
is a symmetric function of $\Gamma$, i.e. $\chi_{\Gamma}^{(B)}=\chi_{-\Gamma}^{(B)}$.
Inset. Scaling of $\chi_{\Gamma}^{(B)}$ approaching the EP from the
right (red line) and from the left (blue). Both susceptibilities were
fitted with the function $\chi_{\Gamma}^{(B)\,\pm}=B_{\Gamma}^{\pm}|\Gamma-\Gamma^{*}|^{-2}$,
with $B_{\Gamma}^{\pm}=0.064$. The second peak of $\chi_{\Gamma}^{(B)\,-}$
corresponds to the EP at $\Gamma=-\hbar\,\Omega_{0,1/2}/R\,c$ outside
the scaling region of the inset. \label{fig:fig4}}
\end{figure}

The biorthogonal fidelity $F_{\Gamma}^{(B)}$ generalizes to the NH
case the distance between two quantum states for different values
of the mass $\Gamma$. In Fig.(\ref{fig:fig4}) we plot $\chi_{\Gamma}^{(B)}$
for the lowest pLL of Eq.(\ref{eq:dirac}) as a function of $\Gamma$.
The susceptibility diverges at the EP and it is symmetric under the
exchange $\Gamma\rightarrow-\Gamma$. In the inset of fig.(\ref{fig:fig4})
we show that $\chi_{\Gamma}^{(B)}$ diverges quadratically approaching
the EP, i.e. $\chi_{\Gamma}^{(B)}\propto|\Gamma-\Gamma^{*}|^{-\gamma,\gamma'}$
with $\gamma=\gamma'=2$ the right/left susceptibility critical exponents.

\begin{figure*}[t]
\includegraphics[width=1\textwidth]{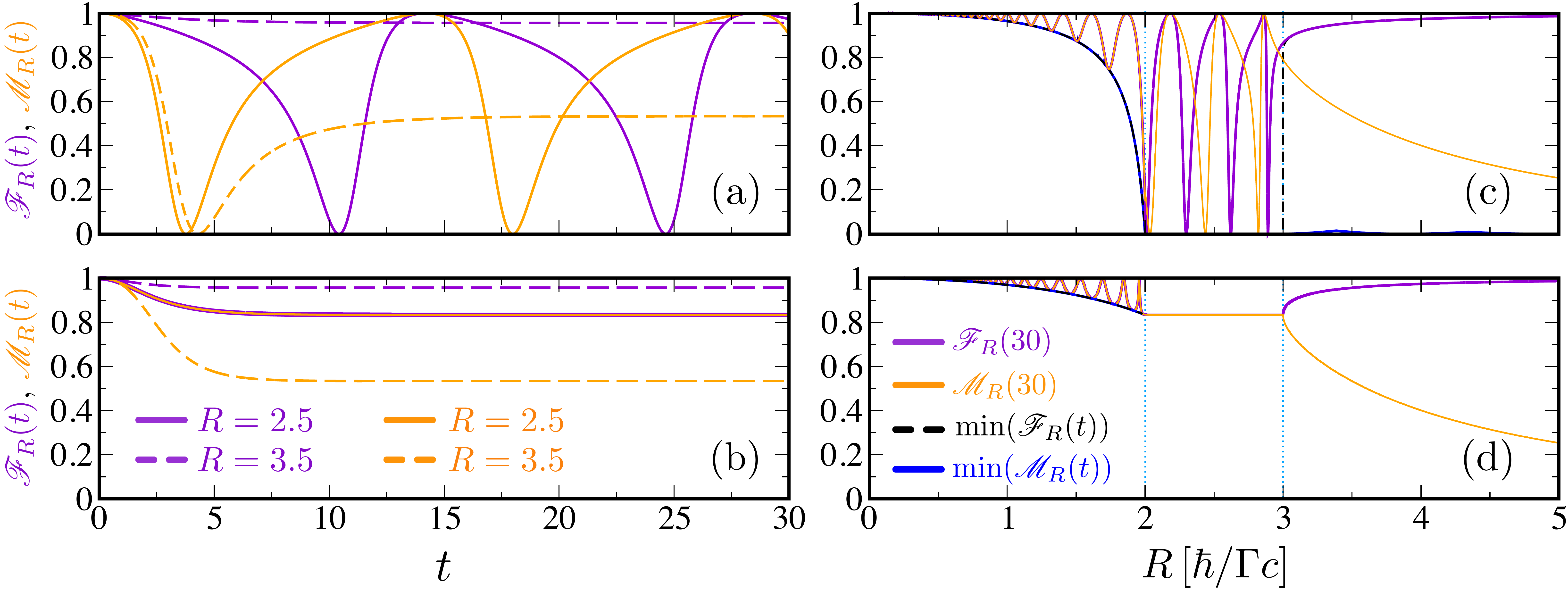} \caption{\emph{Quench dynamics}. Fidelity $\mathcal{F}_{R}(t)$ (purple) and
Loschmidt echo $\mathcal{M}_{R}(t)$ (orange) as a function of time
(a)-(b) and as a function of the radius $R$ (c)-(d) at $t=30$ for
two different mass quenches: (a) and (c) $\Gamma_{i}=1/2$ and $\Gamma_{f}=1/3$;
(b) and (d) $\Gamma_{i}=1/3$ and $\Gamma_{f}=1/2$. For all cases
we choose as the initial state the $|\psi_{0}\rangle=|\phi_{\Gamma_{1},R}^{\left(n,m,\lambda\right)}\rangle$,
the lowest pLL with $(n,m,\lambda)=(0,1/2,1)$ for the Hamiltonian
$\hat{H}_{\Gamma_{i}}$.}
\label{fig:fig3} 
\end{figure*}

\begin{figure}
\begin{centering}
\includegraphics[width=1\columnwidth]{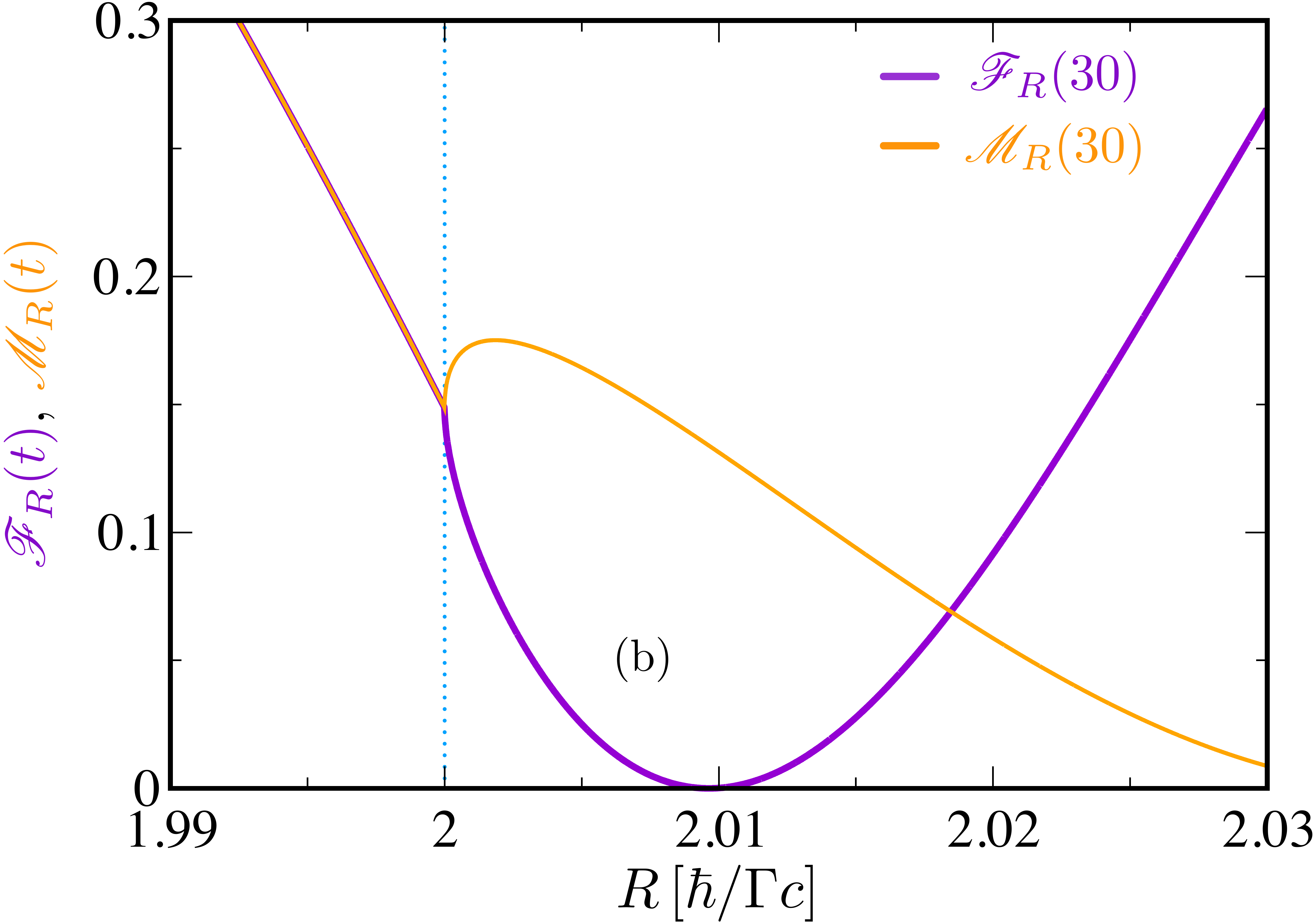} 
\par\end{centering}
\caption{\emph{Quench dynamics close to an EP}. Zoom of Fig.(\ref{fig:fig3})c
for the fidelity $\mathcal{F}_{R}(t)$ (purple) and Loschmidt echo
$\mathcal{M}_{R}(t)$ (orange) as a function of radius $R$ at $t=30$
for the mass quench with $\Gamma_{i}=1/2$ and $\Gamma_{f}=1/3$ close
to an exceptional radius $R=2$. For $R<2$, $\mathcal{F}_{R}(t)=\mathcal{M}_{R}(t)$.
At $R=2$ they bifurcate and assume different values.}
\label{fig:fig6} 
\end{figure}

\subsection{Quench dynamics}

We investigate the quantum evolution after a mass quench in the proximity
of EPs. We monitor the effect on the dynamics of a rapid variation
of the mass from $\Gamma_{i}$ to $\Gamma_{f}$ via the (time-dependent)
fidelity $\mathcal{F}_{R}(t)$ and the Loschimdt echo $\mathcal{M}_{R}(t)$
\citep{longhi,PhysRevA.94.010102}. The Loschmidt echo is a measure
of the irreversibility suffered by the system during its evolution
and generated by differences between forward and backward dynamics
\citep{Goussev2016, Zanardi1,Zanardi2,Moore,Jafari,Mera1}. In contrast,
the fidelity quantifies the deviation of the dynamics of a state induced
by a perturbation. In NH systems, these perturbative effects are strongly
increased in the proximity of the exceptional points \citep{tzeng2020hunting}.
We consider the case where the initial state, $|\psi_{0}\rangle=|\phi_{\Gamma_{1},R}^{\left(n,m,\lambda\right)}\rangle$,
is an eigenstate of $\hat{H}_{\Gamma_{1}}$, the case of a generic
state can be obtained by an immediate generalization of this formalism.
To simplify the notation used here, we omit the indexes $n$ and $m$.
With this choice, the eigenvector is written as $|\phi_{\Gamma_{1},R}^{\left(\lambda\right)}\rangle$,
where the upper index is related to the eigenvalue signal. The Loschmidt
echo is defined as \citep{longhi} 
\begin{equation}
\mathcal{M}_{R}\left(t\right)=\frac{\left|\langle\psi_{0}|\psi_{f}(t)\rangle\right|^{2}}{\langle\psi_{0}|\psi_{0}\rangle\langle\psi_{f}(t)|\psi_{f}(t)\rangle}.\label{eq:loschmidt}
\end{equation}
The initial state is propagated forward in time with the Hamiltonian
$\hat{H}_{\Gamma_{1}}$ and then time reversed and propagated backward
in time with $\hat{H}_{\Gamma_{2}}$ to obtain $|\psi_{f}(t)\rangle$.
Exploiting the orthogonality relations among the eigenstates of the
Dirac Hamiltonian for different masses $\Gamma_{1,2}$ (see Appendix
\ref{sec:Biorthogonal}) the expression of the time-evolved state
can be greatly simplified 
\begin{align}
|\psi_{f}(t)\rangle & =\sum_{\tilde{\lambda}=\pm}e^{i\frac{t}{\hbar}\left[E_{\Gamma_{2},R}^{\left(\lambda\right)}-E_{\Gamma_{1},R}^{\left(\lambda\right)}\right]}|\psi_{\Gamma_{2},R}^{\left(\tilde{\lambda}\right)}\rangle\times\nonumber \\
 & \times\langle\chi_{\Gamma_{2},R}^{\left(\tilde{\lambda}\right)}|\phi_{\Gamma_{1},R}^{\left(\lambda\right)}\rangle.
\end{align}

Similarly, we compute the \emph{time-dependent} fidelity $\mathcal{F}_{R}\{|\psi_{1}^{(\lambda)}(t)\rangle,|\psi_{2}^{(\lambda)}(t)\rangle\}$
defined in Eq.(\ref{eq:fidelity}) between the time-evolved initial
state $|\phi_{\Gamma_{1},R}^{\left(\lambda\right)}\rangle$ under
$\hat{H}_{\Gamma_{1}}$, i.e 
\begin{equation}
|\psi_{1}^{\left(\lambda\right)}\left(t\right)\rangle=e^{-i\frac{t}{\hbar}E_{\Gamma_{1},R}^{\left(\lambda\right)}}|\phi_{\Gamma_{1},R}^{\left(\lambda\right)}\rangle,
\end{equation}
and $\hat{H}_{\Gamma_{2}}$, i.e. 
\begin{equation}
|\psi_{2}^{\left(\lambda\right)}\left(t\right)\rangle=\sum_{\ell=\pm}e^{-i\frac{t}{\hbar}E_{\Gamma_{2},R}^{\left(\lambda\right)}}|\psi_{\Gamma_{2},R}^{\left(\ell\right)}\rangle\langle\chi_{\Gamma_{2},R}^{\left(\ell\right)}|\phi_{\Gamma_{1},R}^{\left(\lambda\right)}\rangle.
\end{equation}
The Loschmidt echo and the fidelity evaluated on the eigenstates of
Eq.(\ref{eq:dirac}) satisfy the relation $\mathcal{F}_{R}^{\left(\lambda\right)}\left(t\right)=\mathcal{M}_{R}^{\left(-\lambda\right)}\left(t\right)$
(see Appendix \ref{sec:Quantum-dynamics-after}). Also, similarly
to the Hermitian case, $\mathcal{F}_{R}^{\left(\lambda\right)}\left(t\right)=\mathcal{M}_{R}^{\left(\lambda\right)}\left(t\right)$
in the TR-invariant phase, whereas they generally differ in the TR-broken
phase. In Fig.(\ref{fig:fig3}) we plot the fidelity and the Loschmidt
echo as a function of time and the radius $R$ of the sphere for quenches
with $\Gamma_{i}>\Gamma_{f}$ (figs.(a) and (c)) and $\Gamma_{i}<\Gamma_{f}$
(figs.(b) and (d)) with the initial state $|\phi_{\Gamma_{1},R}^{(1)}\rangle$,
with $\left(n=0,m=1/2\right)$. For both $\Gamma_{i,f}$ we find two
critical radii $R^{*}=2$ and $R^{*}=3$ (blue dotted lines) at which
we observe singularities for $\text{min}_{t\in[0,\infty[}\mathcal{F}_{R}(t)$
(black dashed lines) and $\text{min}_{t\in[0,\infty[}\mathcal{M}_{R}(t)$
(blue line). For quenches within the TR-invariant region for both
$\hat{H}_{\Gamma_{i}}$ and $\hat{H}_{\Gamma_{f}}$, when $R<2$,
the fidelity and the Loschmidt echo coincides at all times, $\mathcal{F}_{R}(t)=\mathcal{M}_{R}(t)$,
(c-d). When $2<R<3$ they are equivalent only when $\Gamma_{i}<\Gamma_{f}$
and it is independent of the radius, (b) and (d). At larger radii
$R>3$, $\mathcal{F}_{R}(t)>\mathcal{M}_{R}(t)$ for both quenches,
(c) and (d). Notice that in this limit the small minimum value of
the Loschmidt echo $\text{min}_{t\in[0,\infty[}\mathcal{M}_{R}(t)$
in (c) is attained at intermediate times (dashed-orange curve in (a)),
whereas in (d) it coincides with the asymptotic value $\mathcal{M}_{R}(\infty)$
(dashed-orange curve in (b)). In Fig.(\ref{fig:fig3})b we observe
that for large times $\mathcal{F}_{R}(t)$ and $\mathcal{M}_{R}(t)$
converge to a constant. When $\lambda=1$, with $\Gamma_{1}>\Omega_{nm},\ \Gamma_{2}>\Omega_{nm}$
we have asymptotically ($t\rightarrow\infty$) 
\begin{align}
\mathcal{F}_{R}\left(t\rightarrow\infty\right) & =\frac{\left|c_{\Gamma_{2},\Gamma_{1},R}^{\left(1,1\right)}\right|^{2}}{a_{\Gamma_{2},\Gamma_{1},R}^{\left(1,1,1\right)}\langle\psi_{\Gamma_{1},R}^{\left(1\right)}|\psi_{\Gamma_{1},R}^{\left(1\right)}\rangle},\\
\mathcal{M}_{R}\left(t\rightarrow\infty\right) & =\frac{\left|c_{\Gamma_{2},\Gamma_{1},R}^{\left(-1,1\right)}\right|^{2}}{\langle\psi_{\Gamma_{1},R}^{\left(1\right)}|\psi_{\Gamma_{1},R}^{\left(1\right)}\rangle b_{\Gamma_{2},\Gamma_{1},R}^{\left(-1,-1,1\right)}},
\end{align}
where we introduced the constants 
\begin{align}
a_{\Gamma_{2},\Gamma_{1},R}^{\left(q,\ell,\lambda\right)} & =\langle\chi_{\Gamma_{2},R}^{\left(q\right)}|\psi_{\Gamma_{1},R}^{\left(\lambda\right)}\rangle^{\ast}\langle\psi_{\Gamma_{2},R}^{\left(q\right)}|\psi_{\Gamma_{2},R}^{\left(\ell\right)}\rangle\langle\chi_{\Gamma_{2},R}^{\left(\ell\right)}|\psi_{\Gamma_{1},R}^{\left(\lambda\right)}\rangle,\\
b_{\Gamma_{2},\Gamma_{1},R}^{\left(q,\lambda,\ell\right)} & =\langle\chi_{\Gamma_{2},R}^{\left(q\right)}|\psi_{\Gamma_{1},R}^{\left(\lambda\right)}\rangle^{\ast}\langle\chi_{\Gamma_{2},R}^{\left(\ell\right)}|\psi_{\Gamma_{1},R}^{\left(\lambda\right)}\rangle,\\
c_{\Gamma_{2},\Gamma_{1},R}^{\left(\ell,\lambda\right)} & =\langle\chi_{\Gamma_{2},R}^{\left(\ell\right)}|\psi_{\Gamma_{1},R}^{\left(\lambda\right)}\rangle^{\ast}\langle\psi_{\Gamma_{2},R}^{\left(\ell\right)}|\psi_{\Gamma_{1},R}^{\left(\lambda\right)}\rangle.
\end{align}

In Fig.(\ref{fig:fig6}) we zoom on the region around the EP at $R=2$
with $\Gamma_{i}=1/2$ and $\Gamma_{f}=1/3$, where the fidelity and
the Loschmidt echo display a singularity and a bifurcation that display
where the phase transition occurs.

\section{Conclusions}

In this work we studied the influence of geometry in a prototypical
problem of a Dirac particle moving on the surface of a sphere with
an imaginary mass. From the analysis of the spectrum, we found an
infinite sequence of EPs arising from pLLs in the presence of an imaginary
mass. The NH magnetization and fidelity susceptibilities reveal phase
transitions in correspondence of the EPs from TR-invariant to TR-broken
phases. Finally, we investigated the quench dynamics through the Loschmidt
echo and the fidelity and find singular behavior of these quantities
at the EPs. Our work paves the way to study and analyze curvature-dependent
NH phases where the curved background can be experimentally realized
in synthetic systems \citep{Simon,Kollar,Green}. Promising platforms
to implement our model are the recent realizations of fullerene in
a macroscopic-size superconducting resonator \citep{PhysRevLett.115.026801}
and of Dirac surface states in topological nanoparticles \citep{Siroki2016,Rider2020}.
Several branches of future investigation can be anticipated, such
as the influence of hyperbolic geometry \citep{Kollar} and the introduction
of topological defects \citep{deJuan} in NH phases.

\section*{Acknowledgments}

We thank A. Lourenço and R. Arouca for useful discussions, and J.
K. Pachos for valuable comments on the manuscript. This study was
financed in part by the Coordenação de Aperfeiçoamento de Pessoal
de Nível Superior -- Brasil (CAPES) -- Finance Code 001. T.M. acknowledges
CNPq for support through Bolsa de produtividade em Pesquisa n.311079/2015-6.
This work was supported by the Serrapilheira Institute (grant number
Serra-1812-27802), CAPES-NUFFIC project number 88887.156521/2017-00.

\section*{Appendix}

\appendix
this Appendix we present some technical details on the derivation
of the spectrum of the Dirac Hamiltonian with imaginary mass, the
biorthogonal basis, and the relevant scalar products for the dynamics.

\section{\label{sec:Spectrum}Spectrum of the Dirac Hamiltonian with imaginary
mass}

In this Appendix we briefly review the calculations to the eigenvalue
problem for Dirac hamiltonian with imaginary mass on the sphere. First
we shall derive the differential equation for the spinor components,
find the eigenvalues and the general form of the solutions.

The eigenfunctions of $\hat{H}_{\Gamma}$ are two-component spinors
that satisfy the eigenvalue equation given by $\hat{H}_{\Gamma}\phi_{\Gamma,R}^{\left(m\right)}\left(\theta,\varphi\right)=E_{\Gamma,R}^{\left(m\right)}\phi_{\Gamma,R}^{\left(m\right)}\left(\theta,\varphi\right),$
where 
\begin{equation}
\phi_{\Gamma,R}^{\left(m\right)}\left(\theta,\varphi\right)=\frac{e^{im\varphi}}{\sqrt{4\pi}}\left(\begin{array}{c}
\alpha_{\Gamma,R}^{\left(m\right)}\left(\theta\right)\\
\beta_{\Gamma,R}^{\left(m\right)}\left(\theta\right)
\end{array}\right),\label{eq:psi-non-ortho}
\end{equation}
with $m=\pm\frac{1}{2},\pm\frac{3}{2},\ldots$

\begin{subequations}
Now, we have 
\begin{align}
\frac{\hbar c}{R}\hat{h}^{-}\left[e^{im\varphi}\beta_{\Gamma,R}^{\left(m\right)}\left(\theta\right)\right] & =\left(E_{\Gamma,R}^{\left(m\right)}+ic^{2}\Gamma\right)e^{im\varphi}\alpha_{\Gamma,R}^{\left(m\right)}\left(\theta\right);\label{eq:equp}\\
\frac{\hbar c}{R}\hat{h}^{+}\left[e^{im\varphi}\alpha_{\Gamma,R}^{\left(m\right)}\left(\theta\right)\right] & =\left(E_{\Gamma,R}^{\left(m\right)}-ic^{2}\Gamma\right)e^{im\varphi}\beta_{\Gamma,R}^{\left(m\right)}\left(\theta\right).\label{eq:eqdown}
\end{align}
\label{eq:up-down} 
\end{subequations}
 These two equations combine, after a change of variables $\zeta=\cos\theta,$
to give \citep{2002hep.th...12134A,Abrikosov:2001nj} 
\begin{align}
\left[\frac{d}{d\zeta}\left(1-\zeta^{2}\right)\frac{d}{d\zeta}-\frac{m^{2}-\sigma_{z}m\zeta+1/4}{1-\zeta^{2}}\right.\nonumber \\
+\left.\Omega^{2}-\frac{1}{4}\right]\left(\begin{array}{c}
\alpha_{\Gamma,R}^{\left(m\right)}\left(\zeta\right)\\
\beta_{\Gamma,R}^{\left(m\right)}\left(\zeta\right)
\end{array}\right) & =0.
\end{align}
with 
\begin{equation}
\Omega^{2}=\frac{R^{2}}{\hbar^{2}c^{2}}\left(E_{m}^{2}+\Gamma^{2}c^{4}\right).\label{eq:omega}
\end{equation}

Changing the variables as 
\begin{equation}
\left(\begin{array}{c}
\alpha_{m}\left(\zeta\right)\\
\beta_{m}\left(\zeta\right)
\end{array}\right)=\left(\begin{array}{c}
\left(1-\zeta\right)^{\frac{1}{2}\left|m-\frac{1}{2}\right|}\left(1+\zeta\right)^{\frac{1}{2}\left|m+\frac{1}{2}\right|}\xi_{m}\left(\zeta\right)\\
\left(1-\zeta\right)^{\frac{1}{2}\left|m+\frac{1}{2}\right|}\left(1+\zeta\right)^{\frac{1}{2}\left|m-\frac{1}{2}\right|}\eta_{m}\left(\zeta\right)
\end{array}\right),
\end{equation}
one can verify 
\begin{align}
\left\{ \left(1-\zeta^{2}\right)\frac{d^{2}}{d\zeta^{2}}+\left[\frac{m}{\left|m\right|}\sigma_{3}-\left(2\left|m\right|+2\right)\zeta\right]\frac{d}{d\zeta}\right. & +\nonumber \\
-\left.\left|m\right|\left(\left|m\right|+1\right)+\left(\Omega_{nm}^{2}-\frac{1}{4}\right)\right\} \left(\begin{array}{c}
\xi_{m}\left(\zeta\right)\\
\eta_{m}\left(\zeta\right)
\end{array}\right) & =0.
\end{align}
For square integrable solutions \citep{2002hep.th...12134A} on the
interval $x\in\left[-1,1\right]$, we need that $\Omega_{nm}^{2}=\left(n+\left|m\right|+\frac{1}{2}\right)^{2},$
whith non-negative integer $n\geq0.$

Now, $\xi_{nm}\left(\zeta\right)$ and $\eta_{nm}\left(\zeta\right)$
are expressed in terms of the $n$-th order Jacobi polynomial as 
\begin{equation}
\left(\begin{array}{c}
\xi_{nm}\left(\zeta\right)\\
\eta_{nm}\left(\zeta\right)
\end{array}\right)=\left(\begin{array}{c}
c_{nm}^{\alpha}P_{n}^{\left(\left|m-1/2\right|,\left|m+1/2\right|\right)}\left(\zeta\right)\\
c_{nm}^{\beta}P_{n}^{\left(\left|m+1/2\right|,\left|m-1/2\right|\right)}\left(\zeta\right)
\end{array}\right).
\end{equation}
From Eqs.(\ref{eq:up-down}) we find a condition for the coefficients
\begin{subequations}
\begin{align}
c_{\Gamma,R}^{\beta\left(n,m,\lambda\right)} & =-\text{sgn}\left[m\right]\frac{\left[E_{\Gamma,R}^{\left(n,m,\lambda\right)}+ic^{2}\Gamma\right]}{\frac{\hbar c}{R}\Omega_{nm}}c_{\Gamma,R}^{\alpha\left(n,m,\lambda\right)},\\
c_{\Gamma,R}^{\alpha\left(n,m,\lambda\right)} & =-\text{sgn}\left[m\right]\frac{\left[E_{\Gamma,R}^{\left(n,m,\lambda\right)}-ic^{2}\Gamma\right]}{\frac{\hbar c}{R}\Omega_{nm}}c_{\Gamma,R}^{\beta\left(n,m,\lambda\right)}.
\end{align}
\end{subequations}
 The absolute values of the constants can be found from the normalization
conditions, 
\begin{equation}
\int_{0}^{2\pi}\int_{-1}^{1}\left[\phi_{nm}^{\left(\lambda\right)}\left(\zeta,\varphi\right)\right]^{\dagger}\phi_{nm}^{\left(\lambda\right)}\left(\theta,\varphi\right)R^{2}d\zeta d\varphi=1.
\end{equation}
Therefore, 
\begin{equation}
\left|c_{\Gamma,R}^{\alpha\left(n,m,\lambda\right)}\right|=\sqrt{\frac{\frac{\hbar^{2}c^{2}\Omega_{nm}^{2}n!\left(n+2\left|m\right|\right)!}{R^{2}2^{2\left|m\right|-1}\Gamma\left(\Omega_{nm}\right)^{2}R^{2}}}{\frac{\hbar^{2}c^{2}}{R^{2}}\Omega_{nm}^{2}+\left|E_{\Gamma,R}^{\left(n,m,\lambda\right)}+ic^{2}\Gamma\right|^{2}}}.\label{eq:coeficiente}
\end{equation}
The Eq. (\ref{eq:psi-non-ortho}) becomes

\begin{align}
\phi_{\Gamma,R}^{\left(n,m,\lambda\right)}\left(\zeta,\varphi\right) & =c_{\Gamma,R}^{\alpha\left(n,m,\lambda\right)}a_{\Gamma,R}^{\left(n,m,\lambda\right)}\times\nonumber \\
\times & \frac{e^{im\varphi}}{\sqrt{4\pi}}\left(\begin{array}{c}
A_{nm}\left(\zeta\right)\\
-\frac{E_{\Gamma,R}^{\left(n,m,\lambda\right)}+i\Gamma c^{2}}{\text{sgn}\left[m\right]\frac{\hbar c}{R}\Omega_{nm}}B_{nm}\left(\zeta\right)
\end{array}\right),
\end{align}
where 
\begin{subequations}
\begin{align}
A_{nm}\left(\zeta\right) & =\sqrt{1-\zeta}^{\left|m-\frac{1}{2}\right|}\sqrt{1+\zeta}^{\left|m+\frac{1}{2}\right|}P_{n}^{\left(\left|m-\frac{1}{2}\right|,\left|m+\frac{1}{2}\right|\right)}\left(\zeta\right),\\
B_{nm}\left(\zeta\right) & =\sqrt{1-\zeta}^{\left|m+\frac{1}{2}\right|}\sqrt{1+\zeta}^{\left|m-\frac{1}{2}\right|}P_{n}^{\left(\left|m+\frac{1}{2}\right|,\left|m-\frac{1}{2}\right|\right)}\left(\zeta\right),
\end{align}
\end{subequations}
 and $a_{\Gamma,R}^{\left(n,m,\lambda\right)}=\sqrt{E_{\Gamma,R}^{\left(n,m,\lambda\right)}+i\Gamma c^{2}}^{\ast}/\left|\sqrt{E_{\Gamma,R}^{\left(n,m,\lambda\right)}+i\Gamma c^{2}}\right|.$
The introduction of the term $a_{\Gamma,R}^{\left(n,m,\lambda\right)}$
is useful for considerations on the TR symmetry.

\begin{figure*}[t]
\begin{centering}
\includegraphics[width=0.9\textwidth]{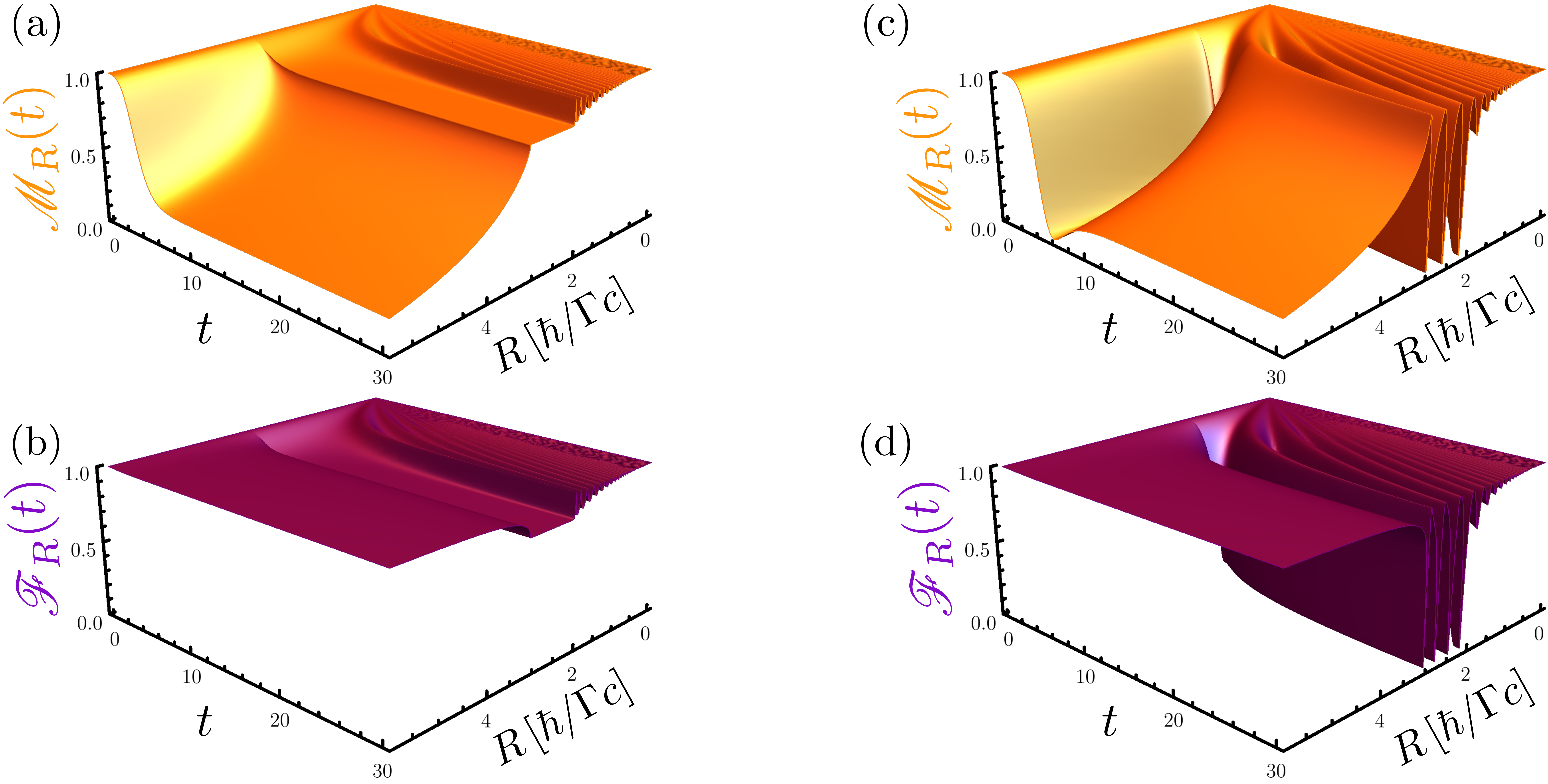} 
\par\end{centering}
\caption{Three-dimensional plot of the fidelity $\mathcal{F}_{R}(t)$ (orange)
and Loschmidt echo $\mathcal{M}_{R}(t)$ (purple) in function of the
sphere radius $R$ and time $t$ for the mass quenches: (a) and (b)
$\Gamma_{i}=1/3$ and $\Gamma_{f}=1/2$. (c) and (d) $\Gamma_{i}=1/2$
and $\Gamma_{f}=1/3$. For all cases we choose as the initial state
the $|\psi_{0}\rangle=|\phi_{\Gamma_{1},R}^{\left(n,m,\lambda\right)}\rangle$,
the lowest pLL with $(n,m,\lambda)=(0,1/2,1)$ for the Hamiltonian
$\hat{H}_{\Gamma_{i}}$}
\label{fig:fig5} 
\end{figure*}

\section{\label{sec:Biorthogonal} Details on the biorthogonal basis: eigenvectors
and scalar products}

In this Appendix we show the eigenvectors of $\hat{H}_{\Gamma}$ and
$\hat{H}_{\Gamma}^{\dagger}$ and the scalar product between then
with different parameters at fixed radius. These relations we will
be useful to study the quantum quenches. To obtain these quantities,
we follow the same steps of Appendix \ref{sec:Biorthogonal} using
Eq.(\ref{eq:bi-ortho-scalar-product}) to impose the correct normalization.

The eigenvectors of the Hamiltonians $\hat{H}_{\Gamma}$ and $\hat{H}_{\Gamma}^{\dagger}$
are given by, respectively,

\begin{equation}
\psi_{\Gamma,R}^{\left(n,m,\lambda\right)}=f_{\Gamma,R}^{\left(n,m,\lambda\right)}\frac{e^{im\varphi}}{\sqrt{4\pi}}\left(\begin{array}{c}
A_{nm}\left(\zeta\right)\\
-\frac{\left[E_{\Gamma,R}^{\left(n,m,\lambda\right)}+i\Gamma c^{2}\right]}{\text{sgn}\left[m\right]\frac{\hbar c}{R}\Omega_{nm}}B_{nm}\left(\zeta\right)
\end{array}\right),\label{eq:psi-sim}
\end{equation}

\begin{equation}
\chi_{\Gamma,R}^{\left(n,m,\lambda\right)}=f_{\Gamma,R}^{\left(n,m,\lambda\right)\ast}\frac{e^{im\varphi}}{\sqrt{4\pi}}\left(\begin{array}{c}
A_{nm}\left(\zeta\right)\\
-\frac{\left[\Xi_{\Gamma,R}^{\left(n,m,\lambda\right)}-i\Gamma c^{2}\right]}{\text{sgn}\left[m\right]\frac{\hbar c}{R}\Omega_{nm}}B_{nm}\left(\zeta\right)
\end{array}\right),\label{eq:chi-sim}
\end{equation}
where 
\begin{equation}
f_{\Gamma,R}^{\left(n,m,\lambda\right)}=\sqrt{\frac{\frac{n!\Gamma\left(n+2\left|m\right|+1\right)\frac{\hbar^{2}c^{2}}{R^{2}}\Omega_{nm}^{2}}{2^{2\left|m\right|-1}\Gamma\left(\Omega_{nm}\right)^{2}R^{2}}}{\frac{\hbar^{2}c^{2}}{R^{2}}\Omega_{nm}^{2}+\left[E_{\Gamma,R}^{\left(n,m,\lambda\right)}+i\Gamma c^{2}\right]^{2}}.}
\end{equation}

The relevant scalar products for different values of $\Gamma$ and
$\lambda$ for the states of Eqs.(\ref{eq:psi-sim}) and (\ref{eq:chi-sim})
are given by 
\begin{align}
\langle\chi_{\Gamma_{1},R}^{\left(\lambda_{1}\right)}|\psi_{\Gamma_{2},R}^{\left(\lambda_{2}\right)}\rangle & =\frac{f_{\Gamma_{1},R}^{\left(n,m,\lambda_{1}\right)}f_{\Gamma_{2},R}^{\left(n,m,\lambda_{2}\right)}R^{2}2^{2\left|m\right|-1}\Gamma\left(\Omega_{nm}\right)^{2}}{\left(\hbar^{2}c^{2}/R^{2}\right)\Omega_{nm}^{2}n!\Gamma\left(n+2\left|m\right|+1\right)}\times\nonumber \\
\times\left(\frac{\hbar^{2}c^{2}}{R^{2}}\Omega_{nm}^{2}\right.+ & \left[E_{\Gamma_{1},R}^{\left(n,m,\lambda_{1}\right)}+i\Gamma_{1}c^{2}\right]\left.\left[E_{\Gamma_{2},R}^{\left(n,m,\lambda_{2}\right)}+i\Gamma_{2}c^{2}\right]\right)
\end{align}
and 
\begin{align}
\langle\psi_{\Gamma_{1},R}^{\left(\lambda_{1}\right)}|\psi_{\Gamma_{2},R}^{\left(\lambda_{2}\right)}\rangle & =\frac{f_{\Gamma_{1},R}^{\left(n,m,\lambda_{1}\right)}f_{\Gamma_{2},R}^{\left(n,m,\lambda_{2}\right)}R^{2}2^{2\left|m\right|-1}\Gamma\left(\Omega_{nm}\right)^{2}}{\left(\hbar^{2}c^{2}/R^{2}\right)\Omega_{nm}^{2}n!\Gamma\left(n+2\left|m\right|+1\right)}\times\nonumber \\
\times\left(\frac{\hbar^{2}c^{2}}{R^{2}}\Omega_{nm}^{2}\right.+ & \left[E_{\Gamma_{1},R}^{\left(n,m,\lambda_{1}\right)}+i\Gamma_{1}c^{2}\right]^{\ast}\left.\left[E_{\Gamma_{2},R}^{\left(n,m,\lambda_{2}\right)}+i\Gamma_{2}c^{2}\right]\right).
\end{align}

\section{\label{sec:Quantum-dynamics-after}Dynamics induced by a mass quench:
Fidelity and Loschmidt echo}

In this Appendix we briefly review the calculations of the Loschmidt
echo and the fidelity. These quantities measure the sensitivity to
perturbations of the backward evolution. 

Given the state $|\psi_{\Gamma_{1},R}^{\left(\lambda\right)}\rangle$
at $t=0$, if $\hat{H}_{\Gamma_{1,2}}$ is independent of time, then
its time evolution is 
\begin{equation}
|\psi_{\Gamma_{1}}^{\left(\lambda\right)}\left(t\right)\rangle=e^{-i\frac{t}{\hbar}\hat{H}_{\Gamma_{1},R}^{\left(\lambda\right)}}|\psi_{\Gamma_{1},R}^{\left(\lambda\right)}\rangle.
\end{equation}
In terms of the biorthogonal basis we can write 
\begin{equation}
|\psi_{\Gamma_{2}}^{\left(\lambda\right)}\left(t\right)\rangle=\sum_{\ell}e^{-i\frac{t}{\hbar}E_{\Gamma_{2},R}^{\left(\lambda\right)}}|\psi_{\Gamma_{2},R}^{\left(\ell\right)}\rangle\langle\chi_{\Gamma_{2},R}^{\left(\ell\right)}|\psi_{\Gamma_{1},R}^{\left(\lambda\right)}\rangle.
\end{equation}
Computing the corresponding bras as 
\begin{align}
\langle\psi_{\Gamma_{1}}^{\left(\lambda\right)}\left(t\right)| & =e^{i\frac{t}{\hbar}E_{\Gamma_{1},R}^{\left(\lambda\right)\ast}}\langle\psi_{\Gamma_{1},R}^{\left(\lambda\right)}|,\\
\langle\psi_{\Gamma_{2}}^{\left(\lambda\right)}\left(t\right)| & =\sum_{\ell}e^{i\frac{t}{\hbar}E_{\Gamma_{2},R}^{\left(\ell\right)\ast}}\langle\chi_{\Gamma_{2},R}^{\left(\ell\right)}|\phi_{\Gamma_{1},R}^{\left(\lambda\right)}\rangle^{\ast}\langle\psi_{\Gamma_{2},R}^{\left(\ell\right)}|,
\end{align}
we define the fidelity $\mathcal{F}_{R}\left(t\right)$ as 
\begin{align}
\mathcal{F}_{R}\left(t\right) & =\frac{\left|\langle\psi_{\Gamma_{1}}^{\left(\lambda\right)}\left(t\right)|\psi_{\Gamma_{2}}^{\left(\lambda\right)}\left(t\right)\rangle\right|^{2}}{\langle\psi_{\Gamma_{1}}^{\left(\lambda\right)}\left(t\right)|\psi_{\Gamma_{1}}^{\left(\lambda\right)}\left(t\right)\rangle\langle\psi_{\Gamma_{2}}^{\left(\lambda\right)}\left(t\right)|\psi_{\Gamma_{2}}^{\left(\lambda\right)}\left(t\right)\rangle},\label{eq:fid}
\end{align}

with 
\begin{align}
\langle\psi_{\Gamma_{1}}^{\left(\lambda\right)}\left(t\right)|\psi_{\Gamma_{2}}^{\left(\lambda\right)}\left(t\right)\rangle & =\sum_{\ell}e^{i\frac{t}{\hbar}\left[E_{\Gamma_{1},R}^{\left(\lambda\right)\ast}-E_{\Gamma_{2},R}^{\left(\ell\right)}\right]}\times\nonumber \\
\times & \langle\psi_{\Gamma_{1},R}^{\left(\lambda\right)}|\psi_{\Gamma_{2},R}^{\left(\ell\right)}\rangle\langle\chi_{\Gamma_{2},R}^{\left(\ell\right)}|\psi_{\Gamma_{1},R}^{\left(\lambda\right)}\rangle,\\
\langle\psi_{\Gamma_{2}}^{\left(\lambda\right)}\left(t\right)|\psi_{\Gamma_{2}}^{\left(\lambda\right)}\left(t\right)\rangle & =\sum_{q}\sum_{\ell}e^{i\frac{t}{\hbar}\left[E_{\Gamma_{2}}^{\left(q\right)\ast}-E_{\Gamma_{2}}^{\left(\ell\right)}\right]}\times\nonumber \\
\times\langle\chi_{\Gamma_{2},R}^{\left(q\right)} & |\psi_{\Gamma_{1},R}^{\left(\lambda\right)}\rangle^{\ast}\langle\psi_{\Gamma_{2},R}^{\left(q\right)}|\psi_{\Gamma_{2},R}^{\left(\ell\right)}\rangle\langle\chi_{\Gamma_{2},R}^{\left(\ell\right)}|\psi_{\Gamma_{1},R}^{\left(\lambda\right)}\rangle,\\
\langle\psi_{\Gamma_{1}}^{\left(\lambda\right)}\left(t\right)|\psi_{\Gamma_{1}}^{\left(\lambda\right)}\left(t\right)\rangle & =e^{i\frac{t}{\hbar}\left[E_{\Gamma_{1}}^{\left(\lambda\right)\ast}-E_{\Gamma_{1}}^{\left(\lambda\right)}\right]}\langle\psi_{\Gamma_{1},R}^{\left(\lambda\right)}|\psi_{\Gamma_{1},R}^{\left(\lambda\right)}\rangle.
\end{align}

Note that we use the definition of $\psi_{\Gamma_{2},R}^{\left(\ell\right)}$
normalized according to the biorthogonal basis, i.e., Eq. (\ref{eq:psi-sim}).

Let us consider an initial state $|\psi_{0}^{\left(\lambda\right)}\rangle$
at time $t=0$ an eigenstate of the $\hat{H}_{\Gamma_{1}},$ i.e.,
$|\psi_{\Gamma_{1},R}^{\left(\lambda\right)}\rangle$. We define the
evolution of this state as $|\psi_{f}^{\left(\lambda\right)}\rangle=e^{i\frac{t}{\hbar}H_{\Gamma_{2}}}e^{-i\frac{t}{\hbar}H_{\Gamma_{1}}}|\psi_{\Gamma_{1}}^{\left(\lambda\right)}\rangle$.
The Loschmidt echo $\mathcal{M}\left(t\right)$ is defined as 
\begin{equation}
\mathcal{M}_{R}\left(t\right)=\frac{\left|\langle\psi_{0}^{\left(\lambda\right)}|\psi_{f}^{\left(\lambda\right)}\rangle\right|^{2}}{\langle\psi_{0}^{\left(\lambda\right)}|\psi_{0}^{\left(\lambda\right)}\rangle\langle\psi_{f}^{\left(\lambda\right)}|\psi_{f}^{\left(\lambda\right)}\rangle},\label{eq:echo}
\end{equation}
with 
\begin{align}
\langle\psi_{0}^{\left(\lambda\right)}|\psi_{f}^{\left(\lambda\right)}\rangle & =e^{-i\frac{t}{\hbar}E_{\Gamma_{1}}^{\left(\lambda\right)}}\sum_{\ell}e^{i\frac{t}{\hbar}E_{\Gamma_{2}}^{\left(\ell\right)}}\langle\psi_{\Gamma_{1}}^{\left(\lambda\right)}|\psi_{\Gamma_{2}}^{\left(\ell\right)}\rangle\langle\chi_{\Gamma_{2}}^{\left(\ell\right)}|\psi_{\Gamma_{1}}^{\left(\lambda\right)}\rangle,\\
\langle\psi_{f}^{\left(\lambda\right)}|\psi_{f}^{\left(\lambda\right)}\rangle & =e^{i\frac{t}{\hbar}\left[E_{\Gamma_{1}}^{\left(\lambda\right)\ast}-E_{\Gamma_{1}}^{\left(\lambda\right)}\right]}\sum_{q}\sum_{\ell}e^{i\frac{t}{\hbar}\left[E_{\Gamma_{2}}^{\left(\ell\right)}-E_{\Gamma_{2}}^{\left(q\right)\ast}\right]}\times\nonumber \\
\times & \langle\chi_{\Gamma_{2}}^{\left(q\right)}|\psi_{\Gamma_{1}}^{\left(\lambda\right)}\rangle^{\ast}\langle\psi_{\Gamma_{2}}^{\left(q\right)}|\psi_{\Gamma_{2}}^{\left(\ell\right)}\rangle\langle\chi_{\Gamma_{2}}^{\left(\ell\right)}|\psi_{\Gamma_{1}}^{\left(\lambda\right)}\rangle,\\
\langle\psi_{0}^{\left(\lambda\right)}|\psi_{0}^{\left(\lambda\right)}\rangle & =\langle\psi_{\Gamma_{1},R}^{\left(\lambda\right)}|\psi_{\Gamma_{1},R}^{\left(\lambda\right)}\rangle.
\end{align}

In Fig.(\ref{fig:fig5}) we show the three-dimensional plot of $\mathcal{F}_{R}(t)$
and $\mathcal{M}_{R}(t)$ for the two quenches with different values
of $\Gamma$.

 \bibliographystyle{apsrev4-2}
\bibliography{bose}

\end{document}